%% file: date2017.tex
\documentclass[10pt,times,mathptm,psfig,twocolumn]{IEEEtran}
\usepackage{algorithmicx}
\usepackage{algorithm}
\usepackage{algpseudocode}
\usepackage{multirow}
\usepackage{subcaption}
\usepackage{setspace}
\usepackage[utf8]{inputenc}
\usepackage{amsmath}
\usepackage{amsfonts}
\usepackage{amssymb}
\usepackage{graphicx}
\usepackage{color}
\begin{document}

\newcommand{\RNum}[1]{\uppercase\expandafter{\romannumeral #1\relax}}

{\small\title{Reverse Engineering of Irreducible Polynomials in GF($2^m$) Arithmetic}}

\author{Cunxi Yu, ~Daniel Holcomb, ~Maciej Ciesielski \\
ECE Department,~University of Massachusetts, Amherst, USA \\
    ycunxi@umass.edu, ~holcomb@engin.umass.edu, ~ciesiel@ecs.umass.edu}

\maketitle

\input{abstract}
{ \\ }

\input{introduction}

\input{background}

\input{preliminaries}

\input{results}

{\textbf{Acknowledgment:} The authors would like to thank Prof. Kalla, University of Utah, for his valuable discussion and the benchmarks; and Dr. Arnaud Tisserand, University Rennes 1 ENSSAT, for his valuable discussion. This work is funded by NSF grants, CCF-1319496 and CCF-1617708.}
\tiny
\vspace{-1mm}
\bibliographystyle{IEEEtran}
\bibliography{verification_ycunxi}
\end{document}

%% file: abstract.tex
{\it\bf Abstract -} 
Current techniques for formally verifying circuits implemented in Galois field ($GF$) arithmetic are limited to those with a known irreducible polynomial $P(x)$. This paper presents a computer algebra based technique that extracts the irreducible polynomial $P(x)$ used in the implementation of a multiplier in GF($2^m$). The method is based on first extracting a unique polynomial in Galois field of each output bit independently. $P(x)$ is then obtained by analyzing the algebraic expression in GF($2^m$) of each output bit. We demonstrate that this method is able to reverse engineer the irreducible polynomial of an \textit{n}-bit $GF$ multiplier in \textit{n} threads. Experiments were performed on \textit{Mastrovito} and \textit{Montgomery} multipliers with different $P(x)$, including NIST-recommended polynomials and optimal polynomials for different microprocessor architectures.

%% file: introduction.tex
\section{Introduction}

Galois field (GF) arithmetic is used to implement critical arithmetic components in communication and security-related hardware. It has been extensively applied in many digital signal processing and security applications, such as Elliptic Curve Cryptography (ECC), Advanced Encryption Standard (AES), and others. Multiplication is one of the most heavily used Galois field computations and is a complexity operation. Specifically, in cryptography systems, the size of Galois field circuits can be very large. Therefore, developing a general formal analysis technique of Galois field arithmetic HW/SW implementations becomes critical.  Contemporary formal techniques, such as \textit{Binary Decision Diagrams} (BDDs), \textit{Boolean Satisfiability} (SAT), \textit{Satisfiability Modulo Theories} (SMT), etc., are not efficient to either the verification or reverse engineering of Galois field arithmetic. The limitations of these techniques when applied to Galois field arithmetic have been addressed \cite{kalla:tcad13}. 

The elements in field GF($2^m$) can be represented using polynomial rings. The field of size $m$ is constructed using \textit{irreducible polynomial} $P(x)$, which includes terms with degree $d$ $\in$ [$0, m$] and coefficients in GF(2). For example, $P(x)$=$x^4+x+1$ is an irreducible polynomial in GF($2^4$).  The multiplication in the field is performed modulo $P(x)$. Theoretically, there is a large number of irreducible polynomials available for constructing the field arithmetic operations in GF($2^m$). However, the choice of irreducible polynomial has great impact on the actual implementation of the resulting GF circuits and the performance of field arithmetic operations. The irreducible polynomials differ in the number of bit-level XOR operations. It is believed that, in general, the irreducible polynomial with minimum number of elements gives the best performance \cite{ciet2002short}. However, a later work \cite{scott2007optimal} demonstrates that the best irreducible polynomial from circuit performance point of view varies in different scenarios, and depends on a computer architecture in which it is used, such as ARM vs. Intel-Pentium. In other words, 1) for GF($2^m$) multiplication, each irreducible polynomial results in a unique implementation; and 2) for a fixed field size, there exist many irreducible polynomials that could be used for constructing the field in different applications. This provides the main motivation for this work.

Computer algebra techniques with polynomial representations is believed to offer best solution for analyzing arithmetic circuits \cite{kalla:tcad13}\cite{STABLE:date11}\cite{sayedformal:date-2016}\cite{ciesielski2015verification}. These work address the verification problems of Galois field arithmetic and integer arithmetic implementations, including abstractions \cite{STABLE:date11}\cite{sayedformal:date-2016}\cite{ciesielski2015verification}. The verification problem is typically formulated as proving that the implementation satisfies the specification. This task is accomplished by performing a series of divisions of the specification polynomial $F$ by the implementation polynomials $B$, representing components that implement the circuit. The techniques based on \textit{Gr{\" o}bner Basis} demonstrate that this approach can efficiently transform the verification problem to \textit{membership testing} of the specification polynomial in the ideals \cite{kalla:tcad13}\cite{sayedformal:date-2016}. A different approach to arithmetic verification of synthesized gate-level circuits has been proposed, using algebraic rewriting technique, which transforms polynomial of primary outputs to polynomial of primary inputs \cite{ciesielski2015verification}. The technique proposed in \cite{kalla:tcad13} has been specifically applied to large GF($2^m$) arithmetic circuits. However, the knowledge of irreducible polynomial is essential to verify the implementations. 

Symbolic computer algebra methods have also been used to reverse engineer the word-level operations for GF circuits and integer arithmetic circuits to speed up the verification performance \cite{yu:2016-abstraction}\cite{sayedequivalence}\cite{kalla:dac2014}. In the work of \cite{yu:2016-abstraction}, the authors proposed a original spectral method based on analyzing the internal polynomial expressions during the rewriting procedure. Sayed-Ahmed et al. \cite{sayedequivalence} introduced a reverse engineering technique in Algebraic Combinational Equivalence Checking (ACEC) process using \textit{Gr{\" o}bner Basis} by converting the function into canonical polynomials. However, both techniques are applicable to integer arithmetic only. In \cite{kalla:dac2014}, an abstraction technique is introduced by analyzing the polynomial representation over $GF(2^m)$. However, similarly to \cite{kalla:tcad13}, it is limited to the implementation with a known irreducible polynomial. In this work, we present a method that is able to reverse engineer the design by extracting the irreducible polynomial $P(x)$ of the GF($2^m$) multiplier, regardless of the GF($2^m$) algorithm used (e.g. $Mastrovito$ and $Montgomery$). This procedure automatically checks the equivalence between the implementation with a golden implementation constructed using the extracted irreducible polynomial $P(x)$.


%% file: background.tex
\section{Background} \label{sec:background}
Different variants of canonical, graph-based representations have been proposed for arithmetic circuit verification, including Binary Decision Diagrams (BDDs) \cite{bryant:1986-bdd}, Binary Moment Diagrams (BMDs) \cite{bmd95}, Taylor Expansion Diagrams (TED) \cite{ted:tcomp06}, and other hybrid diagrams. 
While the canonical diagrams have been used extensively in logic synthesis, high-level synthesis and verification, their application to verify large arithmetic circuits remains limited by the prohibitively high memory requirement of complex arithmetic circuits \cite{ciesielski2015verification}\cite{kalla:tcad13}.  Alternatively, arithmetic verification problems can be modeled and solved using Boolean satisfiability (SAT) or satisfiability modulo theories (SMT). However, it has been demonstrated that these techniques cannot efficiently solve the verification problem of large arithmetic circuits \cite{kalla:tcad13} \cite{cunxi:2016-tcad-verification}. Popular in industry are Theorem Provers, user-driven deductive systems for proving that an implementation satisfies the specification, using mathematical reasoning. However, Theorem Provers require manual guidance and in-depth domain knowledge, which makes it difficult to be applied automatically.

\subsection{Computer Algebra Approaches}

The most advanced techniques that have potential to solve the arithmetic verification problems are those based on symbolic Computer Algebra. 
These methods model the arithmetic circuit specification and its hardware implementation as polynomials  \cite{kalla:tcad13}\cite{STABLE:date11}\cite{ciesielski2015verification}\cite{kalla:dac2014}\cite{wienand:cav08}. 
The verification goal is to prove that implementation satisfies the specification by performing a series of divisions of the specification polynomial $F$ by the implementation polynomials $B=\{f_1, \dots, f_s\}$, representing components that implement the circuit. 
%
The polynomials $f_1,...,f_s$ are called the bases, or {\it generators}, of the ideal $J$. 
Given a set $f_1,...,f_s$ of generators of $J$, 
a set of all simultaneous solutions to a system of equations $f_1$=0; ...,$f_s$=0 is called a {\it variety} $V(J)$. 
Verification problem is then formulated as testing if the specification $F$ vanishes on $V(J)$. 
In some cases, the test can be simplified to testing if $F \in J$, which is known in computer algebra as {\it ideal membership} testing \cite{kalla:tcad13}. 

There are two basic techniques to reduce polynomial $F$ modulo $B$. A standard procedure to test if $F \in J$ is to divide polynomial $F$ by the elements of $B$: \{$f_1,...,f_s$\}, one by one. The goal is to cancel, at each iteration, the leading term of $F$ using one of the leading terms of $f_1,...,f_s$.
If the remainder of the division $r$ is 0, then $F$ vanishes on $V(J)$, proving that the implementation satisfies the specification. However, if $ r \ne 0 $, such a conclusion cannot be made: 
$B$ may not be sufficient to reduce $F$ to 0, and yet the circuit may be correct. To check if $F$ is reducible to zero, a {\it canonical} set of generators, $G=\{g_1,...,g_t\}$, called {\it Gr{\" o}bner basis} is needed. This technique has been successfully applied to Galois field arithmetic \cite{kalla:tcad13} and integer arithmetic circuits \cite{sayedformal:date-2016}. 

Verification work of Galois field arithmetic has been presented in \cite{kalla:tcad13} \cite{kalla:dac2014}. These works provide significant improvement compared to other techniques, since their formulations rely on certain simplifying properties in Galois field during polynomial reduction. Specifically, the problem reduces to the ideal membership testing over a larger ideal that includes $J_0 = \langle x^2-x \rangle $ in ${\mathbb{F}}_2$. In this paper, we provide comparison between this technique and our approach.


\subsection{Function Extraction}

\textit{Function extraction} is an arithmetic verification method originally proposed in \cite{ciesielski2015verification} for integer arithmetic circuits, in $\mathbb{Z}_{2^m}$. 
It extracts a unique bit-level polynomial function implemented by the circuit directly from its gate-level implementation. Extraction is done by \textit{backward rewriting}, i.e., transforming the polynomial representing encoding of the primary outputs (called the \textit{output signature}) into a polynomial at the primary inputs (the \textit{input signature}). This technique has been successfully applied to large integer arithmetic circuits, such as 512-bit integer multipliers. However, it cannot be directly applied to large $GF$ multipliers 
because of exponential size of the intermediate number of polynomial terms before cancellations during rewriting. 
Fortunately, arithmetic $GF(2^{m})$ circuits offer an inherent parallelism which can be exploited in backward rewriting.

In the rest of the paper, we first show how to apply such parallel rewriting in $GF(2^{m})$ circuits while avoiding memory explosion experienced in integer arithmetic circuits. Using this approach, we extract the function of each output element in $\mathbb{F}_{2^m}$ and the function is represented in algebraic expression where all variables are Boolean. Finally, we propose a method to reverse engineer the GF designs by extracting the irreducible polynomial $P(x)$ by analyzing these expressions.

\subsection{Galois Field Multiplication}

Galois field (GF) is an algebraic system with a finite number of elements and two main arithmetic operations, addition and multiplication; other operations can be derived from those two \cite{paar2009understanding}. Galois field with $p$ elements is denoted as $GF(p)$. The most widely-used finite fields are \textit{Prime Fields} and \textit{Extension Fields}, and particularly {\it binary extension fields}. Prime field, denoted $GF(p)$, is a finite field consisting of finite number of integers \{$1,2, ....,p-1$\}, where $p$ is a prime number, with additions and multiplication performed \textit{modulo p}. 
Binary extension field, denoted $GF(2^m)$ (or $\mathbb{F}_{2^m}$), is a finite field with $2^m$ elements. Unlike in prime fields, however, the operations in extension fields are not computed \textit{modulo $2^{m}$}. Instead, in one possible representation (called polynomial basis), 
each element  of $GF(2^m)$ is a {\it polynomial ring} with $m$ terms with the coefficients in $GF(2)$.  Addition of field elements is the usual addition of polynomials, with coefficient arithmetic performed modulo 2.  
Multiplication of field elements is performed modulo {\it irreducible polynomial} $P(x)$ of degree $m$ and coefficients in $GF(2)$. The irreducible polynomial $P(x)$ is analog to the prime number $p$ in prime fields $GF(p)$. 
Extension fields are used in many cryptography applications, such as AES and ECC. In this work, we focus on the verification problem of $GF{(2^{m})}$ multipliers.



Two different GF multiplication structures constructed using different irreducible polynomials $P_{1}(x)$ and $P_{2}(x)$, are shown in Figure \ref{fig:4-bit-gf}. The integer multiplication takes two $n$-bit operands as input and generates a $2n$-bit word, where the values computed at lower significant bits are carried through the carry chain all the way to the most significant bit (MSB). In contrast, there is no carry propagation in GF($2^m$) implementations. To represent the result in $GF{(2^{4})}$, the result of the integer multiplication have to be reduced in $GF(2^4)$ to only four output bits. The result of such a reduction is shown in Figure \ref{fig:4-bit-gf}. In GF($2^4$), the input and output operands are represented using polynomials $A(x)$, $B(x)$ and $Z(x)$, where $A(x)$=$\sum_{n=0}^{n=3} a_{n} \cdot x^{n} $, $B(x)$=$\sum_{n=0}^{n=3} b_{n} \cdot x^{n} $, $Z(x)$=$\sum_{n=0}^{n=3} z_{n} \cdot x^{n}$.

The functions of $s_{i}$ ($i$ $\in$ [0, 6]) are  represented using polynomials in $GF(2)$, namely: $s_{0}$=$a_{0}b_{0}$, $s_{1}$=$a_{1}b_{0}$+$a_{0}b_{1}$, up to $s_{6}$=$a_{3}b_{3}$\footnote{For polynomials in $GF(2)$, "+" is computed as modulo 2.}. The outputs $z_{n}$ ($n$ $\in$ [$0, 3$]) are computed modulo the irreducible polynomial $P(x)$. Using $P_{2}(x)$=$x^4$+$x$+1, we obtain : $z_{0}$=$s_{0}$+$s_{4}$, $z_{1}$=$s_{1}$+$s_{4}$+$s_{5}$, $z_{2}$=$a_0$$b_2$+$a_1$$b_1$+$a_2$$b_0$+$a_2$$b_3$+$a_3$$b_2$+$a_3$$b_3$, and $z_{3}$=$a_0$$b_3$+$a_1$$b_2$+$a_2$$b_1$+$a_3$$b_0$+$a_3$$b_3$. In digital circuits, partial products are implemented using {\sc and} gates, and addition modulo 2 is done using {\sc xor} gates. Note that, unlike in integer multiplication, in $GF(2^m)$ circuits there is no carry out to the next bit. For this reason, as we can see in Figure \ref{fig:4-bit-gf}, the function of each output bit can be computed independently of other bits.

\input{date17-gf_example.tex}

\subsection{Irreducible Polynomials}

For constructing the field $GF{(2^{m})}$, the irreducible polynomial can be either a trinomial, $x^m$+$x^a$+1, or a pentanomial $x^m$+$x^a$+$x^b$+$x^c$+1 \cite{nist-recommend}. In \cite{nist-recommend}, it is stated that the pentanomial is chosen as irreducible polynomial only if an irreducible trinomial doesn't exist. In order to obtain efficient GF multiplication algorithm, it is required that $m$ - $a$ $\geq$ $w$. However, the work of \cite{scott2007optimal} demonstrates that the trinomials are not always better than pentanomials. It means that for a given field size, there could be various irreducible polynomials used in different implementations.


An example of constructing $GF(2^4)$ multiplication using two different irreducible polynomials is shown in Figure \ref{fig:4-bit-gf}. We can see that each polynomial corresponds to a unique multiplication. The performance difference can be evaluated by counting the XOR operations in each multiplication. Since the number of AND and XOR operations for generating partial products (variables $s_{i}$ in Fig. \ref{fig:4-bit-gf}) is always the same, the difference is only caused by the reduction of the corresponding polynomials modulo $P(x)$. The number of XOR operations in reduction process can be counted as the number of terms in each column minus one. For example, the number of XORs using $P_1(x)$ is 3+1+2+3=9; and using $P_2(x)$, the number of XORs is 1+2+2+1=6.


As will be shown in the next section, given the structure of the $GF(2^m)$ multiplication, such as shown in Figure \ref{fig:4-bit-gf}, one can immediatelly identify the irreducible polynomial $P(x)$. This can be done by extracting the terms $s^k$ corresponding to the entry $s^m$ (here $s^4$) in the table and generating the irreducible polynomial beyond $x^m$. 
We know that $P(x)$ must contain $x^m$, and the remaining terms $x^k$ are obtained from the non-zero terms corresponding to the entry $s^m$. 
For the irreducible polynomial $P_1(x)=x^4+x^3+x^0$, the terms $x^3$ and $x^0$ are obtained by noticing the placement of $s^4$ in columns $z_3$ and $z_0$. Similarly, for $P_2(x)=x^4+x^1+x^0$, the terms $x^1$ and $x^0$ are obtained by noticing that $s^4$ is placed in columns $z_1$ and $z_0$. 
The reason for it and the details of this procedure will be explained in the next section.


%% file: date17-gf_example.tex
\begin{figure}[ht]
\small
\centering
\begin{tabular}{lllllll}
             &              &              & $a_3$        & $a_2$        & $a_1$        & $a_0$        \\
             &              &              & $b_3$        & $b_2$        & $b_1$        & $b_0$        \\ \hline
             &              &              & $a_{3}b_{0}$ & $a_{2}b_{0}$ & $a_{1}b_{0}$ & $a_{0}b_{0}$ \\
             &              & $a_{3}b_{1}$ & $a_{2}b_{1}$ & $a_{1}b_{1}$ & $a_{0}b_{1}$ &              \\
             & $a_{3}b_{2}$ & $a_{2}b_{2}$ & $a_{1}b_{2}$ & $a_{0}b_{2}$ &              &              \\
$a_{3}b_{3}$ & $a_{2}b_{3}$ & $a_{1}b_{3}$ & $a_{0}b_{3}$ &              &              &              \\ \hline
$s_6$        & $s_5$        & $s_4$        & $s_3$        & $s_2$        & $s_1$        & $s_0$       
\end{tabular}
\vspace{3mm}
\label{my-label}
        \begin{minipage}{.5\linewidth}
      \centering
\begin{tabular}{cccc}
\multicolumn{4}{l}{$P(x)_{1}$=$x^{4}+x^{3}+1$}                                                                            \\
\multicolumn{1}{c|}{$s_3$} & \multicolumn{1}{c|}{$s_2$} & \multicolumn{1}{c|}{$s_1$} & $s_0$                     \\
\multicolumn{1}{c|}{$s_4$} & \multicolumn{1}{c|}{0}  & \multicolumn{1}{c|}{0} & $s_4$                     \\
\multicolumn{1}{c|}{$s_5$} & \multicolumn{1}{c|}{0}  & \multicolumn{1}{c|}{$s_5$}  & $s_5$                     \\
\multicolumn{1}{c|}{$s_6$} & \multicolumn{1}{c|}{$s_6$} & \multicolumn{1}{c|}{$s_6$}  & $s_6$                      \\ \hline
\multicolumn{1}{l|}{$z_3$} & \multicolumn{1}{l|}{$z_2$} & \multicolumn{1}{l|}{$z_1$} & \multicolumn{1}{l}{$z_0$}
\end{tabular}
    \end{minipage}%
    \begin{minipage}{.5\linewidth}
      \centering
\begin{tabular}{cccc}
\multicolumn{4}{l}{$P(x)_{2}$=$x^{4}+x+1$}                                                                            \\
\multicolumn{1}{c|}{$s_3$} & \multicolumn{1}{c|}{$s_2$} & \multicolumn{1}{c|}{$s_1$} & $s_0$                     \\
\multicolumn{1}{c|}{0} & \multicolumn{1}{c|}{0}  & \multicolumn{1}{c|}{$s_4$} & $s_4$                     \\
\multicolumn{1}{c|}{0} & \multicolumn{1}{c|}{$s_5$}  & \multicolumn{1}{c|}{$s_5$}  & 0                     \\
\multicolumn{1}{c|}{$s_6$} & \multicolumn{1}{c|}{$s_6$} & \multicolumn{1}{c|}{0}  & 0                      \\ \hline
\multicolumn{1}{l|}{$z_3$} & \multicolumn{1}{l|}{$z_2$} & \multicolumn{1}{l|}{$z_1$} & \multicolumn{1}{l}{$z_0$}
\end{tabular}    \end{minipage} 
\caption{\small Two GF($2^4$) multiplications constructed using $P(x)_{1}$=$x^{4}+x^{3}+1$ and $P(x)_{2}$=$x^{4}+x+1$.}
\vspace{-7mm}
\label{fig:4-bit-gf}
\end{figure}

%% file: preliminaries.tex
\section{Approach} \label{sec:preliminaries}

\subsection{Computer Algebraic model}

In this approach, the circuit is modeled as a network of logic elements, including: basic logic gates (AND, OR, XOR, INV), and complex standard cell gates (AOI, OAI, etc.) obtained by synthesis and technology mapping. The following algebraic equations are used to describe basic logic gates in $GF(2^{m})$ \cite{kalla:tcad13}:

{\scriptsize
\vspace{-4mm}
\begin{equation}
     \begin{aligned}
      \text{~~} &\\
       & \neg a = 1 + a ~ mod ~2\\
       & a \wedge b = a\cdot b ~mod~ 2\\
       & a \vee b = a + b + a\cdot b ~mod~ 2\\
       & a \oplus b = a + b ~mod~ 2
     \end{aligned}
\label{eq:boolean-poly}
\end{equation}
}
\vspace{-5mm}
\subsection{Outline of the Approach}

Similarly to the work of \cite{ciesielski2015verification}, the computed function of the circuits is specified by two polynomials, referred to as {\it output signature} and {\it input signature}.
The \textit{output signature} of a $GF(2^{m})$ multiplier is defined as $Sig_{out} = \sum _{i=0} ^{m-1} z_i x^i$, and $z_i \in GF(2)$. The \textit{input signature} of a $GF(2^{m})$ multiplier is $Sig_{in}$ = $\sum _{i=0} ^{m-1} \mathbb{P}_i x^i$, with coefficients (product terms) $\mathbb{P}_i \in GF(2)$, and addition operation performed modulo 2.
As discussed in Section \ref{sec:background} and shown in Figure \ref{fig:4-bit-gf}, given an irreducible polynomial $P(x)$, the input signature $Sig_{in}$ can be computed easily in $GF(2^m)$. The goal of verification is first to transform the output signature, $Sig_{out}$, using polynomial representation of the internal logic elements, into $Sig_{in}$ and then check if $Sig_{in}$ = $Sig_{out}$. 
The following theorem is adopted from \cite{ciesielski2015verification}, where it was initially applied to integer arithmetic circuits in $\mathbb{Z}_{2^m}$. 

\textbf{Theorem 1 (Correctness): } \textit{Given a combinational $GF(2^m)$ arithmetic circuit, composed of logic gates, described by polynomial expressions (Eq. 1), the input signature $Sig_{in}$ computed by backward rewriting is unique and correctly represents the function implemented by the circuit in $GF(2^m)$.}

\textbf{Proof:} The proof relies on the fact that each transformation step (rewriting iteration) is correct. That is, each internal signal is represented by an algebraic expression, which always evaluates to a {\it correct value} in $GF(2^{m})$. This is guaranteed by the correctness of the algebraic model in Eq. (\ref{eq:boolean-poly}), which can be proved by inspection. 
The correctness of the computed signature can be proved by induction on $i$, the step of transforming polynomial $F_i$ into $F_{i+1}$. 
Assuming that $F_{0}$=$Sig_{out}$, and each $F_{i} \in GF(2^{m})$, it is easy to show that  $F_{i+1}$ remains in $GF(2^{m})$, where each variable in $F_{i}$ represents output of some logic gate. During the rewriting process, this variable is substituted by a corresponding polynomial in $GF(2^{m})$. Hence, the resulting polynomial $F_{i+1}$ correctly represents the function $F_{i+1}$ $\in$ $GF(2^{m})$. 
Proof of the uniqueness of the computed signature follows the same reasoning.  
\hfill $\square$

\begin{algorithm}
\scriptsize
\caption{Backward Rewriting in $GF(2^{m})$}\label{alg:commonlogic}
\textbf{Input: Gate-level netlist of $GF(2^{m})$ multiplier}\\ 
\textbf{Input: Output signature $Sig_{out}$} \\
\textbf{Output: algebraic expression of given $Sig_{out}$}
\begin{algorithmic}[1]
\State $\mathcal{P}$=\{$p_{0},p_{1},...,p_{n}$\}: polynomials representing gate-level netlist
\State $F_{0}$=$Sig_{out}$
\For{each polynomial $p_{i}$ $\in \mathcal{P}$} 
\For{output variable $v$ of $p_{i}$ in $F_{i}$}
\State replace every variable $v$ in $F_{i}$ by the expression of $p_{i}$
\State $F_{i}$ $\rightarrow$ $F_{i+1}$
\For{each element/monomial $M$ in $F_{i+1}$}
\If {the coefficient of $M$\%2==0 \\~~~~~~~~~~~~~\textbf{or} $M$ is constant, $M$\%2==0}
\State remove $M$ from $F_{i+1}$
\EndIf
\EndFor
\EndFor
\EndFor \\
\Return $F_{n}$
\end{algorithmic}
\end{algorithm}

The rewriting process is described in {\bf Algorithm 1}.
During the rewriting, the polynomial is simplified 
by applying mod 2 reduction to all its terms.
This is unlike in $\mathbb{Z}_{2^m}$ case, where some terms (with opposite signs) would cancel each other. 

The rewriting algorithm takes the gate-level netlist of a $GF(2^{m})$ circuit as input and first converts each logic gate into equations using Eq. (1). The rewriting process starts with $F_{0}=Sig_{out}$, proceeds in a topological order of the netlist, and ends when all the variables in $F_{i}$ are all primary inputs.  Each iteration includes two steps: Step 1) (lines 4-6 of the Algorithm) substitute the variable of the gate output using the expression in the inputs of the gate (Eq.1), and name the new expression $F_{i+1}$ ; Step 2) (line 4 and lines 8-10) simplify the new expression by removing all the monomials and constants that evaluate to 0 in $GF(2)$. The algorithm outputs the function of the design in $GF(2^m)$ after $n$ iterations, where $n$ is the number of gates in the netlist. 

In addition to verifying the design by comparing the computed polynomial $F_n$ with $Sig_{out}$, the expressions of $F_{n}$ will be used to extract the irreducible polynomial and perform the verification. 

An important observation is that the cancellations of polynomial terms take place only within the expression associated with the same degree of polynomial ring ($Sig_{out}$ is a polynomial ring). In other words, the cancellations resulting the polynomial reduction happen in a logic cone of every output bit independently of other bits, regardless of logic sharing between the cones.

\textbf{Theorem 2 (Parallelizability):} \textit{Given a $GF(2^{m})$ multiplier with $Sig_{out}$ = $F_{0}$ = $z_{0}x^{0}$ + $z_{1}x^{1}$ + ... + $z_{m}x^{m}$; and $F_{i}$=$E_{0}x^{0}$ + $E_{1}x^{1}$ + ... + $E_{m}x^{m}$, where $E_{i}$ is an algebraic expression in $GF(2)$ obtained during rewriting. Then, the polynomial reduction is possible only within a single expression $E_{i}$, for $i$=1, 2, ..., m. }

\textbf{Proof:} Consider a polynomial $E_{i}x^{n_i}$+$E_{k}x^{n_k}$, where $E_{i}$ and $E_{k}$ are simplified in $GF(2)$. That is, $E_{i}=(e^1_{i} + e^2_{i} + ...$), and $E_{k}=(e^1_{k} + e^2_{k} + ...$). After simplifying each of the two polynomials, there are no common monomials between $E_{i}x^{n_i}$ and $E_{k}x^{n_k}$. This is because for any element, $e^l_{i}x^{n_i}$ $\neq$ $e^j_{k}x^{n_k}$, for any pairs of $(i, k)$ and $(l, j)$.
\hfill $\square$

\begin{figure}[t] 
\begin{center}
\includegraphics[scale=0.4]{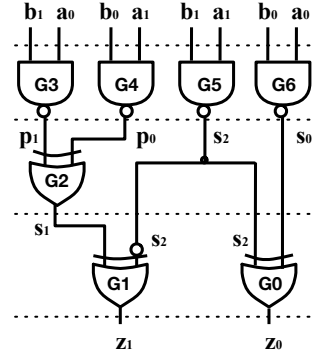}
\caption{\small 2-bit multiplier over $GF(2^2)$ with irreducible polynomial $P(x)=x^{2}+x+1$.}
\vspace{-5mm}
\label{fig:netlist-2bit}
\end{center}
\end{figure}

\input{parallel_example}

{\bf Example 1} (Figure \ref{fig:netlist-2bit}): We illustrate our method using a post-synthesized 2-bit multiplier in $GF(2^2)$, shown in Figure \ref{fig:netlist-2bit}. The irreducible polynomial of this design is $P(x)$ = $x^{2}+x+1$. The goal is to extract algebraic expressions of $z_0$ and $z_1$ by rewriting polynomials from the primary outputs to primary inputs, which is done in parallel ($z_0$ and $z_1$ are rewritten in two threads). The first two transformations rewrite G0 and G1. After this, $z_0$ is rewritten to $s_0$+$s_2$, and $z_1$ is rewritten to $s_1$x+$s_2$x+x. In the rewriting process, we can see that the polynomial reduction happen when there are monomials that are not in GF(2). For example, during the $4^{th}$ iteration of rewriting $z_1$, monomial \textit{2x} is eliminated. Also, we can see that the reductions happen only within the logic cone of each output bit, as proved in Theorem 2. 

In the following, the \textit{out-filed products} are the products $a_i$$b_j$, such that $i+j$$\geq$m. Since these products are associated with bits $s_{i+j}$, they are reduced by $P(x)$.  



\textbf{Theorem 3:} \textit{Given a multiplication in GF($2^m$), let the first out-field product set be $\mathbb{P}_{m}$. Then, the irreducible polynomial $P(x)$ includes $x^m$ , and $x^i$ iff all products in set $\mathbb{P}_{m}$ exist in the algebraic expression of the $i^{th}$ output bits, where $i$ $\leq$ $m$.}

\textbf{Proof:} Based on the definition of field arithmetic, the polynomial basis representation of $\mathbb{P}_{m}$ is $\mathbb{P}_{m}$$x^m$. To reduce $\mathbb{P}_{m}$ into elements in the range [0, $m-1$] (with $m$ output bits), the field reductions are performed modulo irreducible polynomial $P(x)$ with highest degree of $m$. Based on the definition of irreducible polynomial, $P(x)$ is either a trinomial or a pentanomial with degree of $m$. Let $P(x)$ be $x^m$+$P'(x)$. Then, 
{\small
\[
\mathbb{P}_{m}x^m ~mod~ (x^m+P'(x))~=~\mathbb{P}_{m}P'(x)
\]
}
Hence, if $x^i$ exists in $P'(x)$, it also exists in $P(x)$.

\hfill $\square$

{\bf Example 2} (Figure \ref{fig:netlist-2bit}): We illustrate the method of reverse engineering the irreducible polynomial using the 2-bit multiplier in $GF(2^2)$, shown in Figure \ref{fig:netlist-2bit}. The algorithm is shown in Algorithm 2. Using the rewriting technique (Algorithm 1) based on Theorem 1 and 2, we can extract the algebraic expressions of $z_0$=$a_0$$b_0$, and $z_1$x = ($a_0$$b_1$+$a_1$$b_0$+$a_1$$b_1$)x, hence $z_1$=$a_0$$b_1$+$a_1$$b_0$+$a_1$$b_1$ (lines 3 - 5). In this example, $m$=2, hence $\mathbb{P}_{3}$=\{$a_1$$b_1$\}. We can see that both expressions of $z_0$ and $z_1$ include $\mathbb{P}_{3}$, which means that $x^0$ and $x^1$ are included in the irreducible polynomial of this design (lines 6 - 7). Based on Theorem 4, we know that $x^m$ is always included (line 2). Hence, irreducible polynomial of this design is $P(x)$=$x^2+x+1$ ($m$=2) (line 10).


\begin{algorithm}
\scriptsize
\caption{Extracting irreducible polynomial in $GF(2^{m})$}\label{alg:commonlogic}
\textbf{Input: Gate-level netlist/equations of $GF(2^{m})$ multiplier}\\ 
\textbf{Output: Irreducible polynomial $P(x)$}
\begin{algorithmic}[1]
\State $\mathbb{P}_{m}$=\{$a_{m-1}$$b_{1}$, $a_{m-2}$$b_{2}$, ..., $a_{1}$$b_{m-1}$\}
\State $P(x)$=$x^m$: initialize irreducible polynomial
\For{each output bit $z_{i}$} 
\State{apply Algorithm1 \textit{Backward\_rewrite}(netlist/equations, $z_{i}$)} 
\State $EXP_{i}$ $\leftarrow$ \textit{Backward\_rewrite}(netlist, $z_{i}$)
\If {$\mathbb{P}_{m}$ exits in $EXP_i$}
\State $P(x)$~+=~$x^i$
\EndIf
\EndFor \\
\Return $P(x)$
\end{algorithmic}
\end{algorithm}

%% file: parallel_example.tex
\begin{figure}[!htb]
\scriptsize
\centering
\begin{tabular}{|l|c|l|c|}
\hline
\multicolumn{4}{|c|}{ $sig_{out}$=$z_0$+$x$$z_1$} \\ \hline
$Sig_{out0}$=$z_0$ & \begin{tabular}[c]{@{}c@{}}elim\end{tabular} & $Sig_{out1}$=x$\cdot$$z_1$ & \begin{tabular}[c]{@{}c@{}}elim\end{tabular} \\ \hline
G0: $s_0$+$s_2$ & \textit{-} & G0: $z_1$$x$ & - \\ \hline
G1: $s_0$+$s_2$ & \textit{-} & G1: ($s_1$+1+$s_2$)x & - \\ \hline
G2: $s_0$+$s_2$ & \textit{-} & G2: ($p_0$+$p_1$+$s_2$)x+x & - \\ \hline
G3: $s_0$+$s_2$ & \textit{-} & G3: (\textbf{1}+$a_0$$b_1$+$p_0$+$s_2$)x+\textbf{x} & 2x \\ \hline
G4: $s_0$+$s_2$ & \textit{-} & G4: ($a_0$$b_1$+1+$a_1$$b_0$+$s_2$)x & - \\ \hline
G5: $s_0$+1+$a_1$$b_1$ & \textit{-} & G5: ($a_0$$b_1$+$a_1$$b_0$+\textbf{1}+$a_1$$b_1$)x+\textbf{x} & 2x \\ \hline
G6: \textbf{1}+$a_0$$b_0$+$a_1$$b_1$+\textbf{1} & \textit{2} & G6: x($a_1$$b_1$+$a_1$$b_0$+$a_0$$b_1$)+2x & 2x \\ \hline
\multicolumn{4}{|l|}{$z_0$=$a_0$$b_0$+$a_1$$b_1$, $z_1$=$a_1$$b_1$+$a_1$$b_0$+$a_0$$b_1$} \\ \hline
\end{tabular}
\caption{\small Extracting the algebraic expression of $z_0$ and $z_{1}$ in Figure \ref{fig:netlist-2bit}.}
\vspace{-3mm}
\label{fig:parallel}
\end{figure}

%% file: results.tex
\vspace{-4mm}
\section{Results}

The technique described in this paper was implemented in C++. It reverse engineers the irreducible polynomials of GF($2^m$) multiplications by analyzing the algebraic expressions of each element. 
The program was tested on a number of combinational gate-level $GF(2^{m})$ multipliers with different irreducible polynomials including Montgomery multipliers and Mastrovito multipliers. The multiplier generators are taken from \cite{kalla:tcad13}. 
It shows that our technique can successfully reverse engineer the irreducible polynomials of various designs, regardless of the GF($2^m$) algorithm. The experiments were conducted on a PC with Intel(R) Xeon CPU E5-2420 v2 2.20 GHz x12 with 32 GB memory.


We first evaluate our approach using Montgomery and Mastrovito multipliers that are implemented using NIST-recommended irreducible polynomials \cite{nist-recommend}. The experimental results of Mastrovito multipliers with bit-width varying from 64 to 571 bits is shown in Table \ref{tbl:mas} and results of Montgomery multipliers with bit-width varying from 64 to 283 bits is shown in Table \ref{tbl:mont}. Note that we use the flattened version Montgomery multipliers, i.e. we have no knowledge of the block boundaries. The bit-width $m$ of the GF($2^m$) multiplier is shown in the first column. The irreducible polynomials used for constructing those multipliers are shown in the second column. The number of equations that represent the implementation is in the third column; it is also the number of iterations of extracting the polynomial expression of each output bit. 

Our program takes the netlist/equations of the GF($2^m$) implementations, and the number of threads as inputs. Hence, the users can adjust the parallel effort depending on the hardware resource. In this work, all results are performed in 16 threads. The results in Table \ref{tbl:mas} and Table \ref{tbl:mont} show that the proposed technique can extract the irreducible polynomial $P(x)$ of large multipliers, regardless of the GF algorithm.

\input{mas_mult_results.tex}
\vspace{-4mm}
\input{mont_mult_results.tex}

We also apply our technique in the bit-optimized multipliers (Table \ref{tbl:syn}). The multipliers are optimized and mapped using synthesis tool ABC \cite{abc-link}. Comparing Table \ref{tbl:syn} with Tables \ref{tbl:mas} and \ref{tbl:mont}, we can see that it takes much less runtime and memory to extract the irreducible polynomials of the bit-optimized multipliers rather than the non-optimized multipliers. This is because the GF multipliers are implemented without carry chain. As long as the logic cone of each output bit can be reduced, the complexity of extracting the polynomial expressions becomes easier.

\input{synthesis.tex}


One observations is that in Table \ref{tbl:mont}, extracting $P(x)$ of GF($2^{163}$) multiplier requires four times runtime of extracting $P(x)$ of GF($2^{233}$) multiplier. The reason is that the complexity of the GF multiplication using different irreducible polynomials can be very different. The results shown in Table \ref{tbl:diff_p} compare the performance of extracting the irreducible polynomials of GF($2^{233}$) Mastrovito multipliers for different $P(x)$. Those multipliers are implemented with the polynomials shown in Table \ref{tbl:diff_p}, which are optimal irreducible polynomials for different computer architectures \cite{scott2007optimal}. We can see that the runtime varies from 233 seconds to 546 seconds, and memory usage varies from 4.8 GB to 11.7 GB. This is because, for different $P(x)$, the total number of XOR operations can be very different, e.g. as for the GF($2^4$) multiplications discussed in Section \RNum{2}-D. 

\input{other_px.tex}

\begin{figure}[!htb] 
\begin{center}
\includegraphics[scale=0.6]{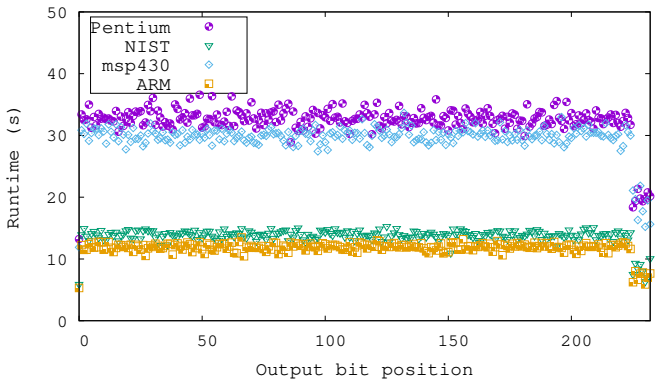}
\caption{\small Runtime of extracting polynomial expressions of each output bit of GF($2^{233}$) multipliers included in Table \ref{tbl:diff_p}.}
\vspace{-8mm}
\label{fig:analysis}
\end{center}
\end{figure}

The complexity of extracting irreducible polynomial is evaluated using the runtime of extracting polynomial expression of each output bit, and finding $\mathbb{P}_{m}$ (Algorithm 2). The analysis results shown in Figure \ref{fig:analysis} are based on the GF($2^{233}$) multipliers used in Table \ref{tbl:diff_p}. The x-axis represents the output bit position, and the y-axis shows the runtime of extracting polynomial expression and finding $\mathbb{P}_{m}$. 

\vspace{-2mm}
\section{Conclusion}
This paper presents a computer algebra based technique that extracts the irreducible polynomial used in the implementation of a multiplier with a given GF($2^m$). The method is based on analyzing the unique polynomial expressions of the output bits in Galois field. The experimental results show that our technique is able to extract the irreducible polynomial up to 571-bit GF multipliers, regardless of the implementation. We analyze the runtime complexity using various irreducible polynomials. 

%% file: mas_mult_results.tex
\begin{table}[!htb]
\scriptsize
\centering
\begin{tabular}{|r|r|r|r|r|}
\hline
\multirow{2}{*}{bit-width $m$} & \multicolumn{1}{c|}{\multirow{2}{*}{Irreducible polynomial P(x)}} & \multirow{2}{*}{\# eqns} & \multicolumn{2}{l|}{Extraction in 16 threads} \\ \cline{4-5} 
 & \multicolumn{1}{c|}{} &  & Runtime(s) & Mem \\ \hline
64 & $x^{64}$+$x^{21}$+$x^{19}$+$x^4$+1 & 21,814 & 9.2  & 37 MB \\ \hline
96 & $x^{96}$+$x^{44}$+$x^{7}$+$x^2$+1 & 51,412 &  13.4  & 86 MB \\ \hline
163 & $x^{163}$+$x^{80}$+$x^{47}$+$x^9$+1 & 153,245 & 158.9  & 253 MB \\ \hline
233 & $x^{233}$+$x^{74}$+1 & 167,803 & 244.9  & 1.5 GB \\ \hline
283 & $x^{283}$+$x^{12}$+$x^{7}$+$x^5$+1 & 399,688 & 704.5 & 4.5 GB \\ \hline
409 & $x^{409}$+$x^{87}$+1 & 508,507 & 1324.7 & 8.3 GB \\ \hline
571 & $x^{571}$+$x^{10}$+$x^{5}$+$x^2$+1 & 1628,170 & 4089.9 & 27.1 GB \\ \hline
\end{tabular}
\caption{\small Results of reverse engineering irreducible polynomials of Mastrovito multipliers using NIST-recommended polynomials.}
\vspace{-3mm}
\label{tbl:mas}
\end{table}

%% file: mont_mult_results.tex
\begin{table}[!htb]
\scriptsize
\centering
\begin{tabular}{|r|r|r|r|r|}
\hline
\multirow{2}{*}{bit-width $m$} & \multicolumn{1}{c|}{\multirow{2}{*}{Irreducible polynomial P(x)}} & \multirow{2}{*}{\# eqns} & \multicolumn{2}{l|}{Extraction in 16 threads} \\ \cline{4-5} 
 & \multicolumn{1}{c|}{} &  & Runtime(s) & Mem \\ \hline
64 & $x^{64}$+$x^{21}$+$x^{19}$+$x^4$+1 & 16.898 & 42.2 & 30 MB\\ \hline
96 & $x^{96}$+$x^{44}$+$x^{7}$+$x^2$+1 & 37,634 & 228.2 & 119 MB \\ \hline
163 & $x^{163}$+$x^{80}$+$x^{47}$+$x^9$+1 & 107,582 & 1614.8 & 2.6 GB \\ \hline
233 & $x^{233}$+$x^{74}$+1 & 219,022 & 461.1 & 4.8 GB \\ \hline
283 & $x^{283}$+$x^{12}$+$x^{7}$+$x^5$+1 & 322,622 & 21520.0 & 7.8 GB \\ \hline
409 & $x^{409}$+$x^{87}$+1 & 672,396 & - & $MO$ \\ \hline
\end{tabular}
\caption{\small Results of reverse engineering irreducible polynomials of Montgomery multipliers using NIST-recommended polynomials. MO=Out of 32 GB}
\label{tbl:mont}
\end{table}

%% file: synthesis.tex
\begin{table}[]
\scriptsize
\centering
\begin{tabular}{|r|r|r|r|r|r|}
\hline
\multirow{2}{*}{$m$} & \multicolumn{1}{c|}{\multirow{2}{*}{Irreducible polynomial}} & \multicolumn{2}{c|}{\textit{Mastrovito-syn}} & \multicolumn{2}{c|}{\textit{Montgomery-syn}} \\ \cline{3-6} 
 & \multicolumn{1}{c|}{} & Runtime(s) & Mem & Runtime(s) & Mem \\ \hline
64 & $x^{64}$+$x^{21}$+$x^{19}$+$x^{4}$+1 & 12.8 & 25 MB & 5.2 & 20 MB \\ \hline
163 & $x^{163}$+$x^{80}$+$x^{47}$+$x^9$+1 & 67.6 & 508 MB & 221.4 & 610 MB \\ \hline
233 & $x^{233}$+$x^{74}$+1 & 149.6 & 1.2 GB & 154.4 & 2.9 GB \\ \hline
409 & $x^{409}$+$x^{87}$+1 & 821.6 & 6.5 GB & 855.4 & 10.3 GB \\ \hline
\end{tabular}
\caption{\small Results of extracting irreducible polynomial of optimized GF($2^m$) Mastrovito and Montgomery multipliers.}
\vspace{-7mm}
\label{tbl:syn}
\end{table}

%% file: other_px.tex
\begin{table}[!htb]
\scriptsize
\centering
\begin{tabular}{|r|r|r|r|}
\hline
\multicolumn{2}{|c|}{Optimal P(x) in GF($2^{233}$)} & Runtime(s) & Mem \\ \hline
\textit{Intel-Pentium} & $x^{233}$+$x^{201}$+$x^{105}$+$x^9$+1 & 546.7 & 11.7 GB \\ \hline
\textit{ARM} & $x^{233}$+$x^{159}$+1 & 233.7 & 5.1 GB \\ \hline
\textit{MSP430} & $x^{233}$+$x^{185}$+$x^{121}$+$x^{105}$+1 & 511.2 & 10.9 GB \\ \hline
\textit{NIST-recommended} & $x^{233}$+$x^{74}$+1 & 244.9 & 4.8 GB \\ \hline
\end{tabular}
\caption{\small Results of extracting irreducible polynomial of GF($2^{233}$) Mastrovito multipliers implemented using different $P(x)$.}
\vspace{-3mm}
\label{tbl:diff_p}
\end{table}

%% file: date2017.bbl
\begin{thebibliography}{10}
\providecommand{\url}[1]{#1}
\csname url@samestyle\endcsname
\providecommand{\newblock}{\relax}
\providecommand{\bibinfo}[2]{#2}
\providecommand{\BIBentrySTDinterwordspacing}{\spaceskip=0pt\relax}
\providecommand{\BIBentryALTinterwordstretchfactor}{4}
\providecommand{\BIBentryALTinterwordspacing}{\spaceskip=\fontdimen2\font plus
\BIBentryALTinterwordstretchfactor\fontdimen3\font minus
  \fontdimen4\font\relax}
\providecommand{\BIBforeignlanguage}[2]{{%
\expandafter\ifx\csname l@#1\endcsname\relax
\typeout{** WARNING: IEEEtran.bst: No hyphenation pattern has been}%
\typeout{** loaded for the language `#1'. Using the pattern for}%
\typeout{** the default language instead.}%
\else
\language=\csname l@#1\endcsname
\fi
#2}}
\providecommand{\BIBdecl}{\relax}
\BIBdecl

\bibitem{kalla:tcad13}
J.~Lv, P.~Kalla, and F.~Enescu, ``{E}fficient {G}robner {B}asis {R}eductions
  for {F}ormal {V}erification of {G}alois {F}ield {A}rithmatic {C}ircuits,''
  \emph{IEEE Trans. on CAD}, vol.~32, no.~9, pp. 1409--1420, September 2013.

\bibitem{ciet2002short}
M.~Ciet, J.-J. Quisquater, and F.~Sica, ``A short note on irreducible
  trinomials in binary fields,'' in \emph{23rd Symposium on Information Theory
  in the BENELUX}, 2002.

\bibitem{scott2007optimal}
M.~Scott, ``Optimal irreducible polynomials for gf (2m) arithmetic.''
  \emph{IACR Cryptology ePrint Archive}, vol. 2007, p. 192, 2007.

\bibitem{STABLE:date11}
E.~Pavlenko, M.~Wedler, D.~Stoffel, W.~Kunz, A.~Dreyer, F.~Seelisch, and
  G.~Greuel, ``Stable: A new qf-bv smt solver for hard verification problems
  combining boolean reasoning with computer algebra,'' in \emph{DATE}, 2011,
  pp. 155--160.

\bibitem{sayedformal:date-2016}
A.~Sayed-Ahmed, D.~Gro{\ss}e, U.~K{\"u}hne, M.~Soeken, and R.~Drechsler,
  ``Formal verification of integer multipliers by combining grobner basis with
  logic reduction,'' in \emph{DATE'16}, 2016, pp. 1--6.

\bibitem{ciesielski2015verification}
M.~Ciesielski, C.~Yu, W.~Brown, D.~Liu, and A.~Rossi, ``{V}erification of
  {G}ate-level {A}rithmetic {C}ircuits by {F}unction {E}xtraction,'' in
  \emph{52nd DAC}.\hskip 1em plus 0.5em minus 0.4em\relax ACM, 2015, pp.
  52--57.

\bibitem{yu:2016-abstraction}
C.~Yu and M.~J. Ciesielski, ``Automatic word-level abstraction of datapath,''
  in \emph{{IEEE} International Symposium on Circuits and Systems, {ISCAS}
  2016, Montr{\'{e}}al, QC, Canada, May 22-25, 2016}, 2016, pp. 1718--1721.

\bibitem{sayedequivalence}
A.~Sayed-Ahmed, D.~Gro{\ss}e, M.~Soeken, and R.~Drechsler, ``Equivalence
  checking using grobner bases,'' \emph{FMCAD'2016}, 2016.

\bibitem{kalla:dac2014}
T.~Pruss, P.~Kalla, and F.~Enescu, ``{E}quivalence {V}erification of {L}arge
  {G}alois {F}ield {A}rithmetic {C}ircuits using {W}ord-{L}evel {A}bstraction
  via {G}r{\"o}bner {B}ases,'' in \emph{DAC'14}, 2014, pp. 1--6.

\bibitem{bryant:1986-bdd}
R.~E. Bryant, ``Graph-based algorithms for boolean function manipulation,''
  \emph{IEEE Trans. on Computers}, vol. 100, no.~8, pp. 677--691, 1986.

\bibitem{bmd95}
R.~E. Bryant and Y.-A. Chen, ``{V}erification of {A}rithmetic {F}unctions with
  {B}inary {M}oment {D}iagrams,'' in \emph{DAC'95}.

\bibitem{ted:tcomp06}
M.~Ciesielski, P.~Kalla, and S.~Askar, ``{T}aylor {E}xpansion {D}iagrams: {A}
  {C}anonical {R}epresentation for {V}erification of {D}ata {F}low {D}esigns,''
  \emph{IEEE Trans. on Computers}, vol.~55, no.~9, pp. 1188--1201, Sept. 2006.

\bibitem{cunxi:2016-tcad-verification}
C.~Yu, W.~Brown, D.~Liu, A.~Rossi, and M.~J. Ciesielski, ``Formal verification
  of arithmetic circuits using function extraction,'' \emph{{IEEE} Trans. on
  {CAD} of Integrated Circuits and Systems}, vol.~35, no.~12, pp. 2131--2142,
  2016.

\bibitem{wienand:cav08}
O.~Wienand, M.~Wedler, D.~Stoffel, W.~Kunz, and G.-M. Greuel, ``{A}n
  {A}lgebraic {A}pproach for {P}roving {D}ata {C}orrectness in {A}rithmetic
  {D}ata {P}aths,'' \emph{CAV}, pp. 473--486, July 2008.

\bibitem{paar2009understanding}
C.~Paar and J.~Pelzl, \emph{Understanding cryptography: a textbook for students
  and practitioners}.\hskip 1em plus 0.5em minus 0.4em\relax Springer Science
  \& Business Media, 2009.

\bibitem{nist-recommend}
NIST, ``Recommended elliptic curves for federal government use,'' 1999.

\bibitem{abc-link}
A.~Mishchenko \emph{et~al.}, ``Abc: A system for sequential synthesis and
  verification,'' \emph{URL http://www. eecs. berkeley. edu/\~{} alanmi/abc},
  2007.

\end{thebibliography}
